\documentclass[twocolumn,aps,prl,groupedaddress,nofootinbib,showpacs]{revtex4}
\usepackage{graphicx}
\usepackage{amsmath}
\usepackage{amssymb}
\usepackage{multirow}
\usepackage{bm}
\usepackage{color}
\usepackage[hypertex]{hyperref}

\usepackage{relsize}
\RequirePackage{xspace}
\usepackage{dcolumn}
\usepackage{bm}

\def\be{\begin{equation}}
\def\ee{\end{equation}}
\def\bd{\begin{displaymath}}
\def\ed{\end{displaymath}}
\def\ba{\begin{eqnarray}}
\def\ea{\end{eqnarray}}

\def\J{\rm J}

\def\L{\rm L}

\def\S{\rm S}

\begin{document}
\title{Search for $\chi_{c_J}(2P)$ from Higher Charmonim E1 Transitions and X,Y,Z States}
\author{
Bai-Qing Li$^a$, Ce Meng$^b$ and Kuang-Ta Chao$^{b,c}$}

\affiliation{$^a$Department of Physics, Huzhou Teachers College,
Huzhou 313000, China;\\$^b$Department of Physics and State Key
Laboratory of Nuclear Physics and Technology, Peking University,
Beijing 100871, China;\\$^c$Center for High Energy physics, Peking
University, Beijing 100871, China}


\begin{abstract}
We calculate the E1 transition widths of higher vector charmonium
states into the spin-triplet 2P states  in three typical potential
models, and discuss the possibility to detect these 2P states via
these E1 transitions. We attempt to clarify the nature of some
recently observed X,Y,Z states by comparing them with these 2P
charmonium states in these E1 transitions. In particular, the
calculated branching ratios of $\psi(4040),\psi(4160)\to
\chi^{'}_{c_J}\gamma$ (J=0,1,2) are found to be in the range of
$10^{-4}\mbox{-}10^{-3}$, and sensitive to the 3S-2D mixing of
$\psi(4040)$ and $\psi(4160)$. The mixing angle may be constrained
by measuring $\psi(4040),\psi(4160)\to Z(3930)\gamma$, if Z(3930) is
identified with the $\chi^{'}_{c_2}$ state, and then be used in
measuring $\chi^{'}_{c_{0,1}}$ states. These processes can be
studied experimentally at $e^+e^-$ colliders such as BEPCII/BESIII
and CESR/CLEO.
\end{abstract}
\pacs{12.39.Jh, 13.20.Gd, 14.40.Pq}

\maketitle

 \section{Introduction}
Since the discovery of $J/\psi$ in
1974~\cite{Aubert:1974js,Augustin:1974xw}, a lot of charmonium
states had been found in the last century.
Among them, the vector states $\psi(4040)$, $\psi(4415)$, and
$\psi(4160)$, which are commonly assigned as $\psi(3S)$, $\psi(4S)$
and $\psi(2 ^3D_1)$ respectively, can be directly produced through
$e^+e^-$ annihilation into one photon, and thus can be readily
detected at $e^+e^-$ colliders like BEPCII/BESIII~\cite{BES2009} and
CESR/CLEO.

Aside from these vector resonances themselves, it is also
interesting to detect the products via E1 transitions of these
vector resonances into lower charmonium states. Especially, the
decay channels $\psi(4040,4160,4415)\rightarrow
\chi^{'}_{c_J}\gamma,\,J=0,1,2$ can be used to detect the 2P
charmonia $\chi^{'}_{c_J}$ and to study the properties of these
missing states. The importance of experimental establishment of
these 2P charmonia consists in at least two aspects. On one side,
the properties of $\chi^{'}_{c_J}$ are important to clarify the
calculations in various potential models and coupled-channel models
(see, e.g.~\cite{Li:2009ad} and references therein). On the other
side, the $\chi^{'}_{c_J}$ could be the candidates of some of the
recently observed charmonium-like states, the so-called "X,Y,Z"
states (for a review see e.g.~\cite{Brambilla:2010cs}).



According to potential model estimates, the spin-triplet 2P states
lie between 3.9 and 4.0 GeV in the charmonium
family~\cite{Eichten:1978tg,Barnes2005,Godfrey:1985xj}.
Experimentally, five charmonium(like) resonances around 3940 MeV
have been found recently. The Z(3930)~\cite{Belle06:Z3930} observed
by the Belle Collaboration in 2006 in the $\gamma\,\gamma$ fusion
experiment with a mass $3929\pm5 \pm2$ MeV is identified with the
$\chi_{c_2}^{'}$. The X(3915), which was also produced in the
$\gamma\,\gamma$ fusion experiment\cite{Uehara:2009tx} and detected
in the $J/\psi\,\omega$ channel  with a mass about 3915 MeV, is
possibly the $\chi_{c_0}^{'}$~\cite{Liu:2009fe}. The Y(3940) and
Y(3915), which were detected in the $B \to J/\psi \omega K$ process
by the Belle Collaboration~\cite{Abe:2004zs} and the BaBar
Collaboration~\cite{Aubert:2007vj} separately are another candidates
for $\chi_{c_{1,0}}^{'}$. The X(3940), which was found by the Belle
Collaboration~\cite{Belle07:X3940} in the recoiling spectrum of
$J/\psi$ in the $e^+e^-$ annihilation process $e^+e^-\to J/\psi +X$
and $e^+e^-\to J/\psi +D\bar{D}^*$, seems not to be a 2P
state~\cite{Li:2009zu,Belle08:X3940&4160}. Another well known state,
the X(3872), which was first found in the $J/\psi~\pi^{+}\pi^{-}$
invariant mass distribution in the B meson decay around 3872
MeV~\cite{Belle0309032} with $J^{PC}=1^{++}$, might be interpreted
as the $D^{*0}D^0+c.c.$ molecule due to the closeness of its mass to
the $D^{*0}D^0$ threshold. But its large production rate in $p-\bar
p$ collisions at the Tevatron and some properties seem to support
that it could be a compact bound state, such as the 2P charmonium
$\chi_{c_{1}}^{'}$, or a mixture of the $\chi_{c_{1}}^{'}$ with the
$D^{*0}D^0+c.c.$ molecule, despite of its lower mass. In fact, the
mass of $\chi_{c_1}(2P)$ can be lowered to below 3.9 GeV if the
color screening effects and coupled channel effects are
considered~\cite{Li:2009zu,Li:2009ad}. So it is interesting to
examine in the E1 transitions of higher charmonia if the X(3872) can
be seen by having the $\chi_{c_1}^{'}$ component in its wave
function.



Because of the phenomenological importance of the $\chi_{cJ}'$
states mentioned above, we will study the production of these states
in the E1 transitions of higher vector charmonia, say, $\psi(4040)$,
$\psi(4160)$ and $\psi(4415)$. The E1 transition width can be
estimated by potential models. Various potential models predict
various transition widths. However, the transition widths of
$\psi(4415)\rightarrow \chi_{c_J}^{'}\,\gamma$ are usually small
because of the smallness of the overlap integral between the wave
functions of 4S and 2P states. On the other hand, the transition
widths of $\psi(4040,4160)\rightarrow \chi_{c_J}^{'}\,\gamma$, are
relatively large (tens to hundreds of KeV as those shown in
Table~\ref{E1rad0} and Table~\ref{E1rad1}) and the corresponding
branching ratios are about $10^{-4}-10^{-3}$. So the
$\chi_{c_J}^{'}$ may  be detected at the upgraded BEPCII/BESIII
through these channels. Note that the E1 transition width depends on
the phase space which is determined by the mass of $\chi_{c_J}^{'}$.
Since the $\chi_{c_2}^{'}$ has been identified with Z(3930), we can
compare various potential model predictions with the experimental
data of $\chi_{c_2}^{'}$ to see if it can be detected in the E1
transitions of higher charmonium states. This comparison may provide
some hints in searching for the other two $\chi_{c_J}^{'}$ states.

We introduce three typical potential models in section II, and
calculate in section III the E1 transition widths of
$\psi(4040,4160,4415)\rightarrow \chi_{c_J}^{'}\,\gamma$
 with both lowest- and
first-order wave functions in the non-relativistic expansion within
these models. We discuss the possibility for producing
$\chi_{c_J}^{'}$ from E1 transitions of higher excited charmonium
states and compare them with those "X,Y,Z" states in section IV,
where the effects of S-D mixing of $\psi(4040)$ and $\psi(4160)$ on
the E1 transition widths are also considered. A summary is given in
section V.


\section{The potential models}
There are many phenomenologically  successful potential models in
the literature. Among them the Cornell
model~\cite{Eichten:1978tg}(here we mark it by Model II) is well
known, which describes the charmonium system quite well. However,
the predicted masses of higher charmonium states seem to be larger
than their experimental values~\cite{Li:2009zu}. A distinct example
is the mass of $\chi_{c_2}^{'}$ which is about 40 MeV larger than
the experimental value. The screened potential model (see
Ref~\cite{Li:2009zu} and  references therein)(here we mark it by
Model I) was proposed to lower the masses of higher charmonia. So we
take it here to estimate the E1 transition production of
$\chi_{c_J}^{'}$. The third model we take was proposed by Ding et
al.~\cite{Ding:1985rh,Ding:1991vu}(here we mark it by Model III), in
which the Coulomb potential has a running coupling constant. The
Hamiltonian in these models have the form of $H
=-\frac{\vec{P}^2}{m_c}+ V_{V}(r) + V_{S}(r)$, where $V_{V}(r)$ is
vector potential and $V_{S}(r)$ is
scalar potential and $m_c$ is the mass of charm quark.%

The potentials in Model I~\cite{Li:2009zu}  are:
\begin{equation}
\label{potential1}
 V_{V}(r) = -\frac{4}{3}\frac{\alpha_C}{r},\quad V_{S}(r) = \lambda (\frac{1-e^{-\mu r}}{\mu}),
\end{equation}
where $\mu$ is the screening factor which makes the long range
scalar part $V_S(r)$ become flat when $r \gg \frac{1}{\mu}$ and
still linearly rising when $r \ll \frac{1}{\mu}$, $\lambda$ is the
linear potential slope, and $\alpha_C$ is the coefficient of the
Coulomb potential. The model parameters are chosen following
Ref~\cite{Li:2009zu}:
\begin{eqnarray}
 \label{para1}
 \alpha_C=0.5007,~~~~~~\lambda=0.21GeV^2, \nonumber\\
\mu=0.0979GeV,~~~~~~~m_c=1.4045GeV,
\end{eqnarray}
where $\alpha_C\approx \alpha_{s}(m_c v_c)$ is essentially the
strong coupling constant at the scale $m_c v_c$. Here $\mu$ is the
characteristic scale for color screening, and $1/{\mu}$ is about 2
$fm$, implying that at distances larger than $1/{\mu}$ the static
color source in the $c\bar c$ system gradually becomes neutralized
by the produced light quark pair, and string breaking emerges.

The potentials in Model II~\cite{Eichten:1978tg} are:
\begin{equation}
\label{potential2}
 V_{V}(r) = -\frac{4}{3}\frac{\alpha_C}{r},\quad V_{S}(r) = \lambda  r,
\end{equation}
with model parameters taken similar to Ref.~\cite{Barnes2005}:
\begin{equation}
 \label{para2}
 \alpha_C=0.5461,\,\lambda=0.1425GeV^2,\, m_c=1.4794GeV \nonumber
\end{equation}

The potentials in Model III~\cite{Ding:1985rh,Ding:1991vu} are
\begin{equation}
\label{potential3}
 V_{V}(r) = \frac{8\pi}{25}\frac{1}{\mbox{ln}(\Lambda r)} \frac{1-\Lambda r}{1+\Lambda r},\,\, V_{S}(r) = \lambda  r\,-\,C,
\end{equation}
with parameters
\begin{eqnarray}
 \label{para3}
 \Lambda=0.47GeV,~~~~~~\lambda=0.22GeV^2, \nonumber\\
  m_c=1.84GeV,~~~~~~~C=-0.975GeV.
\end{eqnarray}

The potentials above are used to calculate the lowest-order and the
first-order non-relativistic wave functions. For the first-order
relativistic corrections to the wave functions, we include the
spin-dependent part of $H_{SS}$, $H_{LS}$, $H_{T}$ and the
spin-independent part $H_{SI}$ as perturbations.

The spin-spin contact hyperfine interaction is
\begin{equation}
\label{Hss1}
 H_{SS} =\frac{2}{3m_c^2}\vec{S_1}\cdot \vec{S_2} \bigtriangledown^2
 V_V.
\end{equation}
For Coulombic vector potential, $\nabla^2\, (\frac{1}{r})\propto
\delta^3(\vec{r})$ which gives too large a mass splitting of
$J/\psi-\eta_c$, so we make a substitution as in
Ref.~\cite{Barnes2005} for Model I and Model II:
\begin{equation}
\label{Hss2}
 H_{SS} =\frac{32\pi\alpha_C}{9 m_c^2}\, \tilde \delta_{\sigma}(r)\, \vec
{S}_c \cdot \vec {S}_{\bar c}\, ,
\end{equation}
where $\tilde \delta_{\sigma}(r) = (\sigma/\sqrt{\pi})^3\,
e^{-\sigma^2 r^2}$ and $\sigma=1.362GeV$ in Model I and
$\sigma=1.0946GeV$ in Model II.

The spin-orbit term is
\begin{equation}
\label{Hls}
 H_{LS} = \frac{1}{2m_{c}^{2}r} (3V_{V}^{'}(r)- V_{S}^{'}(r)) \vec
 {L} \cdot \vec {S},
\end{equation}
and the tensor force term is
\begin{equation}
\label{Ht}
 H_{T} = \frac{1}{12m_{c}^{2}}(\frac{1}{r}V_{V}^{'}(r)-V_{V}^{''}(r)) S_{12},
\end{equation}
where $S_{12}=3(\vec{\sigma_1}\cdot \hat{r})(\vec{\sigma_2}\cdot
\hat{r})-\vec{\sigma_1}\cdot \vec{\sigma_2}$.

The spin-independent part is a bit complicated. We take the form as
Ref.~\cite{Lucha:1991vn}:
\begin{eqnarray}
\label{Hsi}
H_{SI}&=&-\frac{\vec{P}^4}{4m_c^3}+\frac{1}{4m_c^2}\bigtriangledown ^2V_V(r)\nonumber\\
&&-\frac{1}{2m_c^2}\left\{\left\{
\vec{P}_1\cdot V_V(r) \Im \cdot \vec{P}_2 \right\}\right\}\nonumber\\
    &&+\frac{1}{2m_c^2}\left\{\left\{\vec{P}_1\cdot \vec{r}\frac{V_V^{'}(r)}{r}\vec{r}\cdot \vec{P}_2
    \right\}\right\},
\end{eqnarray}
where $\vec{P}_1$ and $\vec{P}_2$ are momenta of c and $\bar c$
quarks in the rest frame of charmonium, respectively, which satisfy
$\vec{P}_1=-\vec{P}_2=\vec{P}$, $\Im$ is the unit second-order
tensor, and $\{\{\quad\}\}$ is the Gromes's notation
\begin{equation}
\{\{\vec{A}\cdot \Re\cdot
\vec{B}\}\}=\frac{1}{4}(\vec{A}\vec{B}:\Re+\vec{A}\cdot \Re
\vec{B}+\vec{B}\cdot \Re \vec{A}+\Re:\vec{A}\vec{B}),
\end{equation}
where $\Re$ is a second-order tensor.

 Note that we do not include
the contributions from the scalar potential in $H_{SI}$ since it is
still unclear how to deal with the spin-independent corrections
arising from the scalar potential theoretically.

\section{E1 transition Widths}

E1 transitions of higher excited S- and D-wave charmonium states are
of interest here because they can be used to produce and identify
P-wave states. For the E1 transition width for charmonium, we use
the formula of Ref.~\cite{Kwong:1988ae}:
\begin{eqnarray}
&&\Gamma_{\rm E1}( {\rm n}\, {}^{2{\S}+1}{\rm L}_{\J} \to {\rm n}'\,
{}^{2{\S}'+1}{\rm L}'_{{\J}'} + \gamma)\nonumber\\
&&=\frac{4}{3}\, C_{fi}\, \delta_{{\S}{\S}'} \, e_c^2 \, \alpha \,
|\,\langle f | \, r \, |\, i \rangle\, |^2 \, {\rm E}_{\gamma}^3 \,
\end{eqnarray}
 where E$_{\gamma}$ is the emitted photon energy.

The spatial matrix element
\begin{equation}
\label{matrix}
<f|r|i>=\int_{0}^{\infty}R_{f}(r)R_{i}(r)r^{3}dr\,,
\end{equation}
involves the initial and final state radial wave functions, and the
angular matrix element $C_{fi}$ is
\begin{equation}
C_{fi}=\hbox{max}({\L},\; {\L}')\; (2{\J}' + 1) \left\{ { {{\L}'
\atop {\J}} {{\J}' \atop {\L}} {{\S} \atop 1}  } \right\}^2 .
\end{equation}

We are only interested in initial states with $J^{PC}=1^{--}$, i.e.,
$\psi(4040)$, $\psi(4415)$  and $\psi(4160)$, since they can be
easily produced in $e^+e^-$ annihilation and can transit into
spin-triplet 2P states by emitting a photon.

For the masses of initial and final states used to calculate
$E_{\gamma}$ in above three models, we take the central values from
PDG(2010)~\cite{Nakamura:2010zzi} if the states are well established
experimentally. The mass of $\chi_{c_0}^{'}$ is supposed to be 3915
MeV, while for the mass of $\chi_{c_1}^{'}$ we choose two different
values: 3872 MeV and 3915 MeV.

The calculated results with lowest-order wave functions are listed
in Table~\ref{E1rad0}. The results of Barnes, et
al.~\cite{Barnes2005}, which are similar to the Model II
and the results of Godfrey, et al.~\cite{Godfrey:1985xj}, with a
relativized  Cornell model, are also listed in Table~\ref{E1rad0}
for comparison. From Table~\ref{E1rad0}, one can see that the widths
$\Gamma(\psi(4415)\to\chi_{cJ}'\gamma)$, are very small due to large
cancelation in the overlap integral between wave functions of 4S and
2P states, and therefore are very sensitive to the model details. On
the other hand, the predictions for the widths
$\Gamma(\psi(4040,4160)\to\chi_{cJ}'\gamma)$ are large and
insensitive to model details, resulting in quite steady values in
different models.

The results obtained with first-order wave functions are listed in
Table~\ref{E1rad1}. From (7-10), one can see that the corrections to
the non-relativistic potential involve some derivative terms, which
make the potential to be more attractive towards the origin. As a
result, the wave functions with relativistic corrections will be
thinner than those without relativistic corrections. This effect
usually reduces the spatial matrix elements $<f|r|i>$ defined in
(\ref{matrix}), which can be seen directly by comparison between the
results of $\Gamma(\psi(4040,4160)\to\chi_{cJ}'\gamma)$ listed in
Table~\ref{E1rad0} and in Table~\ref{E1rad1}. However, the
relativistic corrections can also change the node structures of the
wave functions of higher exited states, such as $\psi(4415)$, and
make the cancelation in the overlap integral between the wave
functions of 4S and 2P states more modest, and this can be seen in
Table~\ref{E1rad1}. On the other hand, the transition width is
proportional to the factor $E_\gamma^3$, which favors initial states
with higher masses. Thus, the decay widths
$\Gamma(\psi(4415)\to\chi_{cJ}'\gamma)$ listed in Table~\ref{E1rad1}
become larger. But we should mention that these results are not very
reliable and are more sensitive to the model details than those of
$\Gamma(\psi(4040,4160)\to\chi_{cJ}'\gamma)$.

We calculate the E1 transition branching ratios of $\psi(4040)$,
$\psi(4160)$, and $\psi(4415)$ with their total width taken from
PDG(2010)~\cite{Nakamura:2010zzi}. Since the errors of the total
widths are relatively small for these states, we only take the
central values of the total widths in calculating the branching
ratios and do not consider the errors.


\begin{table*}
\caption{E1 transition widths and branching ratios of charmonium
states with the lowest-order wave functions in various potential
models. The masses and total widths of the initial states used in
the calculation are the PDG~\cite{Nakamura:2010zzi} central values,
while the masses of the final states in the Model I-III calculations
are denoted by the numbers in the parentheses. For comparison, the
results of Refs.~\cite{Barnes2005} and~\cite{Godfrey:1985xj} are
also listed.} \vskip 0.3cm
\begin{ruledtabular}
\begin{tabular}{ll|lll|lll|lllll|lllll }
\multicolumn{2}{c|}{Process}          &\multicolumn{3}{c|}{$<f|r|i>\,(GeV^{-1}$)} &\multicolumn{3}{c|}{$\rm k\,(MeV)$}                     & \multicolumn{5}{c|}{$\Gamma_{\rm thy}$~(keV)}                                     & \multicolumn{5}{c}{$Br_{\rm thy}~(\times 10^{-4})$}                                  \\
Intial  &Final                    &I     &II    &III                          &Ours  &Ref.\cite{Barnes2005}&Ref.\cite{Godfrey:1985xj}     &I         &II      &III     &Ref.\cite{Barnes2005}&Ref.\cite{Godfrey:1985xj}       &I   &II   &III  &Ref.\cite{Barnes2005}&Ref.\cite{Godfrey:1985xj}      \\
\hline
 $\psi(4040)$&$\chi^{'}_{c_2}(3929)$    &-4.9  &-4.4  &-3.4                         &109  &67                  &119                          &74        &59      &36      &14                  &48                             &9.3 &7.4  &4.5  &1.8                 &6.0\\
             &$\chi^{'}_{c_1}(3872)$    &-4.9  &-4.4  &-3.4                         &164  &113                 &145                          &151       &122     &74      &39                  &43                             &18.9&15.3 &9.3  &4.9                 &5.4\\
             &$\chi^{'}_{c_1}(3915)$    &-4.9  &-4.4  &-3.4                         &122  &113                 &145                          &63        &51      &31      &39                  &43                             &7.9 &6.4  &3.9  &4.9                 &5.4\\
             &$\chi^{'}_{c_0}(3915)$    &-4.9  &-4.4  &-3.4                         &122  &184                 &180                          &21        &17      &10      &54                  &22                             &2.6 &2.1  &1.3  &6.8                 &2.8\\
\hline
 $\psi(4160)$&$\chi^{'}_{c_2}(3929)$    &5.0   &4.6   &3.8                          &218  &183                 &210                          &12        &10      &7.3     &5.9                 &6.3                            &1.2 &0.97 &0.71 &0.57                &0.61  \\
             &$\chi^{'}_{c_1}(3872)$    &5.0   &4.6   &3.8                          &271  &227                 &234                          &355       &299     &210     &168                 &114                            &34.5&29.0 &20.4 &16.3                &11.1\\
             &$\chi^{'}_{c_1}(3915)$    &5.0   &4.6   &3.8                          &231  &227                 &234                          &219       &185     &132     &168                 &114                            &21.3&18.0 &12.8 &16.3                &11.1\\
             &$\chi^{'}_{c_0}(3915)$    &5.0   &4.6   &3.8                          &231  &296                 &269                          &292       &247     &173     &483                 &191                            &28.3&24.0 &16.8 &46.9                &18.5 \\
\hline
 $\psi(4415)$&$\chi^{'}_{c_2}(3929)$    &-0.013&0.093 &-0.028                       &465  &421                 &446                          &0.04      &2.1     &0.2     &0.62                &15                             &0.006&0.34&0.03 &0.1                 &2.4 \\
             &$\chi^{'}_{c_1}(3872)$    &-0.013&0.093 &-0.028                       &515  &423                 &469                          &0.04      &1.7     &0.2     &0.49                &0.92                           &0.006&0.27&0.03 &0.08                &0.15 \\
             &$\chi^{'}_{c_1}(3915)$    &-0.013&0.093 &-0.028                       &477  &423                 &469                          &0.03      &1.4     &0.1     &0.49                &0.92                           &0.005&0.23&0.02 &0.08                &0.15 \\
             &$\chi^{'}_{c_0}(3915)$    &-0.013&0.093 &-0.028                       &477  &527                 &502                          &0.01      &0.45    &0.04    &0.24                &0.39                           &0.002&0.07&0.006&0.04                &0.06\\
\end{tabular}
\end{ruledtabular}
\label{E1rad0}
\end{table*}

\begin{table*}
 \caption{E1 transition widths and branching ratios of charmonium states in
various potential models with the first-order wave functions. The
masses and total widths of the initial states  used in the
calculation are the PDG~\cite{Nakamura:2010zzi} central values,
while the masses of the final states in the Model I-III calculations
are denoted by the numbers in the parentheses.} \vskip 0.3cm
\begin{ruledtabular}
\begin{tabular}{ll|lll| l| lll|lll }
\multicolumn{2}{c|}{Process}  &\multicolumn{3}{c|}{$<f|r|i>\,(GeV^{-1}$)}   &$\rm k\,(MeV)$     & \multicolumn{3}{c|}{$\Gamma_{\rm thy}$~(keV)}                 & \multicolumn{3}{c}{$Br_{\rm thy}~(\times 10^{-4})$}    \\
    Initial  &Final                   &$I^{'}$&$II^{'}$&$III^{'}$            &                   &$I^{'}$ &$II^{'}$ &$III^{'}$        &$I^{'}$ &$II^{'}$ &$III^{'}$  \\
\hline
 $\psi(4040)$&$\chi^{'}_{c_2}(3929)$   &-4.3   &-3.9    &-3.1                 &109               &56      &47       &29               &7.0 &5.9 &3.6\\
             &$\chi^{'}_{c_1}(3872)$   &-3.7   &-3.4    &-2.8                 &164               &88      &72       &50               &11.0&9.0 &6.3\\
             &$\chi^{'}_{c_1}(3915)$   &-3.7   &-3.4    &-2.8                 &122               &37      &30       &21               &4.6 &3.8 &2.6\\
             &$\chi^{'}_{c_0}(3915)$   &-3.0   &-2.7    &-2.5                 &122               &7.9     &6.2      &5.3              &0.99&0.78&0.66\\
\hline
 $\psi(4160)$&$\chi^{'}_{c_2}(3929)$   &4.3    &4.1     &3.4                  &218               &9.2     &8.2      &5.9              &0.89&0.80&0.57 \\
             &$\chi^{'}_{c_1}(3872)$   &3.6    &3.4     &3.2                  &271               &189     &169      &147              &18.3&16.4&14.3  \\
             &$\chi^{'}_{c_1}(3915)$   &3.6    &3.4     &3.2                  &231               &117     &105      &91               &11.4&10.2&8.8  \\
             &$\chi^{'}_{c_0}(3915)$   &2.7    &2.6     &2.9                  &231               &89      &81       &97               &8.6 &7.9 &9.4\\
\hline
 $\psi(4415)$&$\chi^{'}_{c_2}(3929)$   &-0.42  &-0.13   &-0.19                &465               &42      &4.1      &8.7              &6.8&0.66 &1.4 \\
             &$\chi^{'}_{c_1}(3872)$   &-1.1   &-0.77   &-0.39                &515               &219     &116      &30               &35.3&18.7&4.8  \\
             &$\chi^{'}_{c_1}(3915)$   &-1.1   &-0.77   &-0.39                &477               &174     &92       &24               &28.1&14.8&3.9  \\
             &$\chi^{'}_{c_0}(3915)$   &-1.8   &-1.4    &-0.65                &477               &164     &105      &22               &26.5&16.9&3.5   \\
\end{tabular}
\end{ruledtabular}
\label{E1rad1}
\end{table*}

\section{Discussions On XYZ States}

In this section, we fucus on the implication of the results of
$Br(\psi(4040,4160,4415)\to\chi_{cJ}'\gamma)$ on searching for XYZ
states in these channels. The 3S-2D mixing effects of $\psi(4040)$
and $\psi(4160)$ are also considered in details.

\subsection{Z(3930)}
The Z(3930) was found by the Belle
Collaboration~\cite{Belle06:Z3930} in the process $\gamma \gamma
\rightarrow D \bar{D}$ with
\begin{eqnarray}
M(Z(3930))&=&3929 \pm 5 \pm 2~\mbox{MeV},\label{m:Z3930}\\
\Gamma(Z(3930))& = &29 \pm 10\pm 2~\mbox{MeV},\\
\hspace{-5mm}\Gamma_{\gamma\gamma}{\cal B}(Z(3930)\to D\bar{D})&=&
0.18 \pm 0.05 \pm 0.03~\mbox{KeV},\label{Z3930PW}
\end{eqnarray}
and confirmed by the BaBar Collaboration~\cite{:2010hk} with
\begin{eqnarray}
M(Z(3930))&=&3926.7 \pm 2.7 \pm 1.1~\mbox{MeV},\\
\Gamma(Z(3930))& = &21.3 \pm 6.8\pm 3.6~\mbox{MeV},\\
\hspace{-5mm}\Gamma_{\gamma\gamma}{\cal B}(Z(3930)\to D\bar{D})&=&
0.24 \pm 0.05 \pm 0.04~\mbox{KeV}.
\end{eqnarray}

 The production rate and the angular distribution in the
$\gamma\gamma$ center-of-mass frame suggest that this state is the
previously unobserved $\chi_{c_2}^{'}$~\cite{Belle06:Z3930,:2010hk}.
Its mass, however, is about 40-50 MeV larger than the commonly
predicted value in the quenched potential model (see, e.g.
\cite{Barnes2005}). A lower mass can be obtained by considering the
color screening effect described in Model I~\cite{Li:2009zu} in
which the predicted mass is 3937 MeV.

Since $Z(3930)$ is established as the candidate of $\chi_{c_2}^{'}$,
searching for $Z(3930)$ in the E1 transitions of
$\psi(4040,4160,4415)$ is important to further verify this
assignment and can also be a good criterion in searching for and
identifying other $\chi_{cJ}'$ states in these transitions.

From Table~\ref{E1rad0} and Table~\ref{E1rad1}, we can see the
transition width of $\psi(4040)\rightarrow \chi_{c_2}^{'}\, \gamma$
 is 36-74 KeV with the lowest-order wave functions and 29-56 KeV with the first-order
wave functions in our calculations within Models I-III and in
Ref.~\cite{Godfrey:1985xj}. The corresponding branching ratio is
4.5-9.3$\times 10^{-4}$ and 3.6-7.0$\times 10^{-4}$, respectively.
The branching ratio of order of $10^{-4}$ is encouraging to detect
$\chi_{c_2}^{'}$ in $\psi(4040)$ E1 transitions. Note that the
results of Ref.~\cite{Barnes2005} is notably small than our results.
This is mainly because, the mass of $\chi_{c_2}^{'}$ used in
Ref.~\cite{Barnes2005} is larger than ours and the corresponding
energy of the emitted photon is much smaller than ours.

The branching ratio of $\psi(4160)\rightarrow \chi_{c_2}^{'}\,
\gamma$ is about one-fifth of that of $\psi(4040)\rightarrow
\chi_{c_2}^{'}\, \gamma$ but is still close to $1\times 10^{-4}$.
The branching ratio of $\psi(4415)\rightarrow \chi_{c_2}^{'}\,
\gamma$ is sensitive to the model details. So it is difficulty to
predict how large is the branching ratio of $\psi(4415)\rightarrow
\chi_{c_2}^{'}\, \gamma$ in potential models.

\subsection{X(3915),Y(3940),Y(3915)}

The X(3915)~\cite{Uehara:2009tx}, Y(3940)~\cite{Abe:2004zs} and
Y(3915)~\cite{Aubert:2007vj} are all observed in the invariance mass
distribution of $J/\psi\,\omega$ in processes
\begin{eqnarray}
\gamma\,\gamma \rightarrow &X(3915)&\rightarrow J/\psi\,\omega,\\
B\rightarrow K\,Y(3940), &Y(3940)&\rightarrow J/\psi\,\omega,\\
B\rightarrow K\,Y(3915), &Y(3915)&\rightarrow J/\psi\,\omega,
\end{eqnarray}
 with masses
\begin{eqnarray}
M(X(3915))&=&3915 \pm 3 \pm 2~\mbox{MeV},\\
M(Y(3940))&=&3943 \pm 11 \pm 13~\mbox{MeV},\\
M(Y(3915))&=&3914.6^{+3.8}_{-3.4}\pm 2.0~\mbox{MeV},
\end{eqnarray}
total widths
\begin{eqnarray}
\Gamma(X(3915))& = &17 \pm 10\pm 3~\mbox{MeV},\\
\Gamma(Y(3940))& = &87 \pm 22\pm 26~\mbox{MeV},\\
\Gamma(Y(3915))& = &34^{+12}_{-8} \pm 5~\mbox{MeV},
\end{eqnarray}
and partial widths
\begin{widetext}
\begin{eqnarray}
\Gamma_{\gamma\gamma}&\times\,\,\,{\cal B}(X(3915)\to J/\psi\,
\omega)&=\left\{\begin{array}{ll}61\pm17\pm8\quad
\mbox{eV},&J^P=0^+\\18\pm5\pm2\quad \mbox{eV},&J^P=2^+,
\mbox{helicity-2}\label{X3915PW}
\end{array}\right.\\
 {\cal B}(B\rightarrow K\,Y(3940)) &\times\,\,\,{\cal B}(Y(3940)\rightarrow
J/\psi\,\omega)&= (7.1 \pm 1.3 \pm 3.1)\times 10^{-5},\\
 {\cal B}(B^+\rightarrow K^+\,Y(3915)) &\times\,\,\,{\cal B}(Y(3915)\rightarrow
J/\psi\,\omega)&= (3.5 \pm 0.2 \pm 0.4)\times 10^{-4},\\ {\cal
B}(B^0\rightarrow K^0\,Y(3915)) &\times\,\,\,{\cal
B}(Y(3915)\rightarrow J/\psi\,\omega)&= (3.1 \pm 0.6 \pm 0.3)\times
10^{-4}.
\end{eqnarray}
\end{widetext}

Although the differences of masses and total widths  of the three
signals are within $2\sigma$, especially those of X(3915) and
Y(3915) are less than one $\sigma$, it is not clear whether these
three signals come from the same particle.
BaBar~\cite{Aubert:2007vj} considers Y(3915) and Y(3940) as the same
state since the smaller values of both the mass and total width of
Y(3915) derived from fitting data by BaBar can  partially be
attributed to larger data sample used by BaBar, which enable them to
use smaller $J/\psi\,\omega$ mass bin in their
analysis~\cite{Yuan:2009iu}. Y(3940) and X(3915) are also considered
to be the same state in
Refs.~\cite{Uehara:2009tx,Yuan:2009iu,Zupanc:2009qc,Godfrey:2009qe}.


Besides whether they are the same particle or not, there are no
decisive interpretations of these states. That they are detected in
the $J/\psi\,\omega$ channel but not in the $D\bar{D}$ or
$D\bar{D}^{*}$ channel makes people suspect they are not
conventional charmonium states. Ref.~\cite{Uehara:2009tx} argues
that X(3915) is  not an excited charmonium state but favors the
prediction of $D^{*}\bar{D}^{*}$ bound state
model~\cite{Branz:2009yt}. The Y(3915) and Y(3940) are also
interpreted as $D^{*}\bar{D}^{*}$ molecular states
by~\cite{Liu:2009ei,Zhang:2009vs}.

However, Liu et al.~\cite{Liu:2009fe} pointed out that the node
effect of excited wave functions may change the open charm decay
widths dramatically. They calculated the open charm decay of
$\chi_{c_J}^{'}$ and $\chi_{c_J}^{''}$ in the $^3P_0$ model, and
argued that X(3915) is the $\chi_{c_0}^{'}$ state.

If they are conventional charmonia, they are probably the
$\chi_{c_J}^{'}$ states since their masses are consistent with
potential model predictions~\cite{Barnes2005,Godfrey:1985xj}, and
they have positive charge parity since they decay to
$J/\psi\,\omega$.

The X(3915), which is observed in $\gamma\gamma$ fusion, may be
$\chi_{c_0}^{'}$ or $\chi_{c_2}^{'}$. But the mass of X(3915) is
about $2\sigma$ different from that of Z(3930) and from
(\ref{Z3930PW}) and (\ref{X3915PW}). Furthermore, if X(3915)is the
same meson as Z(3930), we would get ${\cal
B}(\chi_{c_2}^{'}\rightarrow J/\psi\,\omega)/{\cal
B}(\chi_{c_2}^{'})\rightarrow D\bar{D})\approx 0.1\pm 0.04$, which
seems to be too large for charmonium. So X(3915) is unlikely to be
$\chi_{c_2}^{'}$ and we tend to regard it as a candidate for
$\chi_{c_0}^{'}$.

The Y(3915), which is produced in B decays, has so close mass and
total width to the X(3915) that we suspect they are the same state.
However, if Y(3915) is not X(3915), then it may be $\chi_{c_1}^{'}$.
Since the E1 transition rate from $\psi(4040)$ to $\chi_{c_1}^{'}$
is three to four times larger than that to $\chi_{c_0}^{'}$ with
both lowest-order and first-order wave functions if they have the
same mass, we can distinguish between them by measuring these E1
transitions.

The Y(3940) is unlikely to be $\chi_{c_1}^{'}$. If it is
$\chi_{c_1}^{'}$, its main decay mode ought to be $D\bar{D}^{*}$.
However, Belle gives ${\cal B}(Y(3940)\rightarrow
J/\psi\,\omega)/{\cal B}(Y(3940)\rightarrow D^0\bar{D}^{*0})> 0.71$
at 90\% CL~\cite{Aushev:2010zz}, which disfavor the $\chi_{c_1}^{'}$
assignment. But Y(3940) might still be a candidate for
$\chi_{c_0}^{'}$.

We may detect and identify them in the E1 transitions of higher
charmonium states. We can see from Table~\ref{E1rad0} and
Table~\ref{E1rad1} that the transition widths are $O(10)$ KeV for
$\psi(4040)\rightarrow \chi_{c_{0,1}}^{'}\,\gamma$ and $O(100)$ KeV
for $\psi(4160)\rightarrow \chi_{c_{0,1}}^{'}\,\gamma$,
corresponding to branching ratios of order $10^{-4}$ and $10^{-3}$,
in all listed potential models. The effect of corrections of wave
functions is moderate for $\psi(4040)$ and $\psi(4160)$ transitions.
The large branching ratios may make the detection of
$\chi_{c_{0,1}}^{'}$ possible.

The transitions of $\psi(4415)\rightarrow
\chi_{c_{0,1}}^{'}\,\gamma$ are model dependent and sensitive to the
node structure, just like $\psi(4415)\rightarrow
\chi_{c_2}^{'}\,\gamma$ as we have remarked.

\subsection{X(3872)}
The $X(3872)$ was first observed by Belle~\cite{Belle0309032} in the
$J/\psi~\pi^{+}\pi^{-}$ invariant mass distribution in
$B^{+}\rightarrow K^{+} J/\psi~\pi^{+}\pi^{-}$ decay as a very
narrow peak ($\Gamma_X<2.3$ MeV) around 3872 MeV. The mass of
$X(3872)$ in the $J/\psi~\pi^+\pi^-$ mode was recently updated by
CDF Collaboration~\cite{CDF08:X3872:mass} as
\be M(X(3872))=3871.61\pm0.16\pm0.19~\mbox{MeV},\label{M:X3872} \ee
which is very close to the $D^0\bar{D}^{*0}$ threshold
$m(D^0\bar{D}^{*0})=3871.81\pm0.36$ MeV~\cite{Cleo07:D-mass}.
Moreover, analyzes both by Belle~\cite{Belle05:X3872:J^PC} and
CDF~\cite{CDF06:X3872:J^PC} favor the quantum number
$J^{PC}=1^{++}$. The mass seems to be too small for a
$J^{PC}=1^{++}$ $\chi_{c_1}^{'}$ charmonium, but the color-screening
effects and the coupled channel effects may lower its mass towards
3872 Mev~\cite{Li:2009zu,Li:2009ad}, as has been declared in Section
I.

%
%
%
There are lots of possible explanations for
X(3872)(see~\cite{Li:2009zu,Yuan:2009iu} for a review). Aside from
the most popular one, i.e., the $D^0\bar{D}^{*0}$ molecular state,
the $1^{++}$
charmonium~\cite{Li:2009zu,Barnes:2003vb,Eichten:2004uh} or a mixed
$1^{++}$ charmonium-$D^0\bar{D}^{*0}$
state~\cite{Meng:2005er,Suzuki:2005ha} for X(3872) was also
proposed. More data samples are needed to distinguish between
various explanations.

Recently, an analysis of the $\omega\to \pi^+\pi^-\pi^0$ spectrum in
the decay $B\to K X\to K J/\psi\omega$ performed by BaBar
\cite{delAmoSanchez:2010jr} claimed that the $J^{PC}$ of $X(3872)$
might disfavor $1^{++}$, as had widely been accepted, but favor
$2^{-+}$. If this result is confirmed, the natural assignment is the
${}^1D_2$ charmonium $\eta_{c2}$. However, the mass of this D-wave
state is about 3.80-3.84 GeV in the potential
models~\cite{Eichten:1978tg,Barnes2005,Godfrey:1985xj}, which is too
low to be the candidate of $X(3872)$. Besides, some recent
theoretical studies on the properties of $\eta_{c2}$ indicate that
it is not apt to the the candidate of
$X(3872)$~\cite{Jia:2010jn,Burns:2010qq,Kalashnikova:2010hv,Fan:2011}.
Therefore, we will ignore this possibility in the following
analysis.

We are interested here in detecting X(3872) in the E1 transitions of
higher charmonia, if X(3872) is the 2P charmonium $\chi_{c1}'$ or
contains some 2P charmonium component in its wave function. One can
see from Table~\ref{E1rad0} that the branching ratio for
$\psi(4040)\rightarrow \chi_{c_1}^{'}\,\gamma$ is $(0.93-1.89)\times
10^{-3}$ and that for $\psi(4160)\rightarrow \chi_{c_1}^{'}\,\gamma$
is $(2.04-3.05)\times 10^{-3}$ in our calculation with zero-order
wave functions. The large branching ratios may enable us to find
$\chi_{c_1}^{'}$ in the $e^+e^-$ machines and compare it with
X(3872). Note that Ref.~\cite{Barnes2005} and
Ref.~\cite{Godfrey:1985xj} give smaller branching ratios. It is
partly because they take a larger mass for $\chi_{c_1}^{'}$ in
calculations. If they take the mass of $\chi_{c_1}^{'}$ to be the
same as us, the differences will diminish. This means the branching
ratios are not sensitive to models. The calculated results with
relativistically corrected wave functions are a bit smaller but
still quite large (see Table~\ref{E1rad1}). It indicates that the
results are not sensitive to the nodes of wave functions and should
be reliable.

\subsection{3S-2D Mixing}
The $\psi(4040)$ and $\psi(4160)$ are commonly assigned as $\psi(3
^3S_1)$  and $\psi(2 ^3D_1)$ respectively. Therefore, the above
results of E1 transitions are all based on these simple assignments.
However, the observed leptonic width $\Gamma^{ee}(4040)\approx
\Gamma^{ee}(4160)$ is inconsistent with this picture. The simplest
explanation is that they are roughly 1:1 mixtures of $\psi(3^3S_1)$
and $\psi(2^3D_1)$. Neither the tensor force nor the coupled channel
effects can cause such strong mixing (see Ref.~\cite{Close:1995eu}
and references therein) and the mixing mechanism remains unknown.
Here we do not consider the mixing mechanism, and simply assume that
they are mixtures of $\psi(3^3S_1)$ and $\psi(2^3D_1)$ with a mixing
angle $\theta$, and calculate how the E1 transition widths vary with
$\theta$. In this case, the $\psi(4040)$ and $\psi(4160)$ are
expressed as
\begin{eqnarray}
&&|\psi(4040)\rangle=|3{}^3{\rm S}_1\rangle \cos\theta+
 |2{}^3{\rm D}_1\rangle\sin\theta,\\
&&|\psi(4160)\rangle=-
 |3{}^3{\rm S}_1\rangle\sin\theta+ |2{}^3{\rm
D}_1\rangle\cos\theta.
\end{eqnarray}
Using the data of leptonic decay widths of $\psi(4040)$ and
$\psi(4160)$~\cite{Nakamura:2010zzi} as inputs, one can determine
the mixing angle $\theta$ in the potential models. It is about
$-35^o$ or $+55^o$ in all the three models used in Sec. II. And this
is consistent with earlier estimates of the mixing
angle\cite{Chao2008,Badalian2009}

\begin{figure}
 \centering
\includegraphics[width=10cm,height=7.5cm]{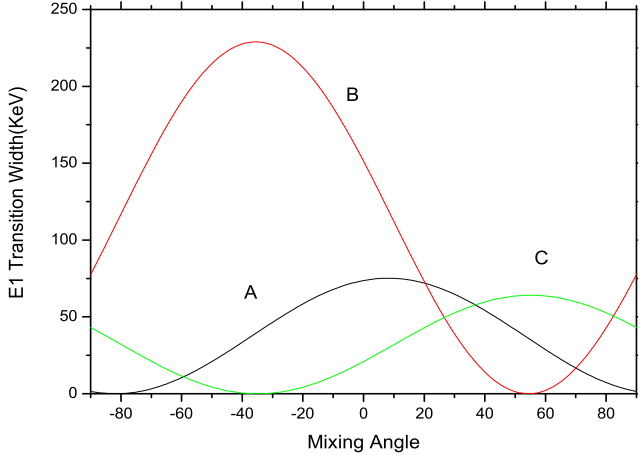}
\caption{E1 transition widths of $\psi(4040)$ varying with 3S-2D
mixing angle in Model I. A: $\psi(4040) \to
\chi^{'}_{c_2}(3929)\gamma$, B: $\psi(4040) \to
\chi^{'}_{c_1}(3872)\gamma$, C: $\psi(4040) \to
\chi^{'}_{c_0}(3915)\gamma$.} \label{ModIpsi4040to2P}
\end{figure}

\begin{figure}
 \centering
\includegraphics[width=10cm,height=7.5cm]{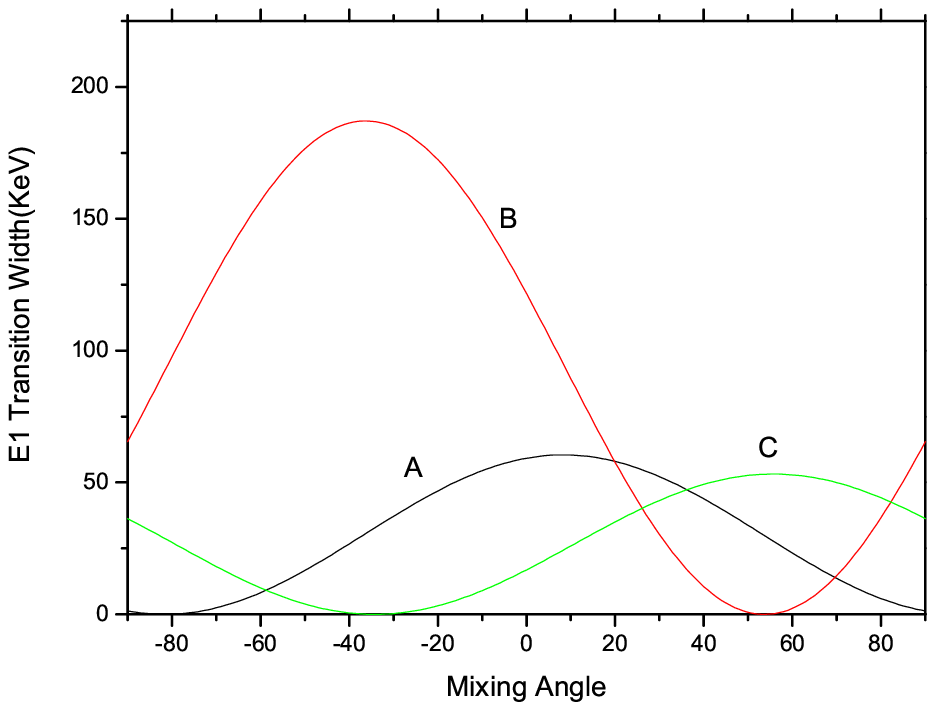}
\caption{E1 transition widths of $\psi(4040)$ varying with 3S-2D
mixing angle in Model II. A: $\psi(4040) \to
\chi^{'}_{c_2}(3929)\gamma$, B: $\psi(4040) \to
\chi^{'}_{c_1}(3872)\gamma$, C: $\psi(4040) \to
\chi^{'}_{c_0}(3915)\gamma$.}   \label{ModIIpsi4040to2P}
\end{figure}

\begin{figure}
 \centering
\includegraphics[width=10cm,height=7.5cm]{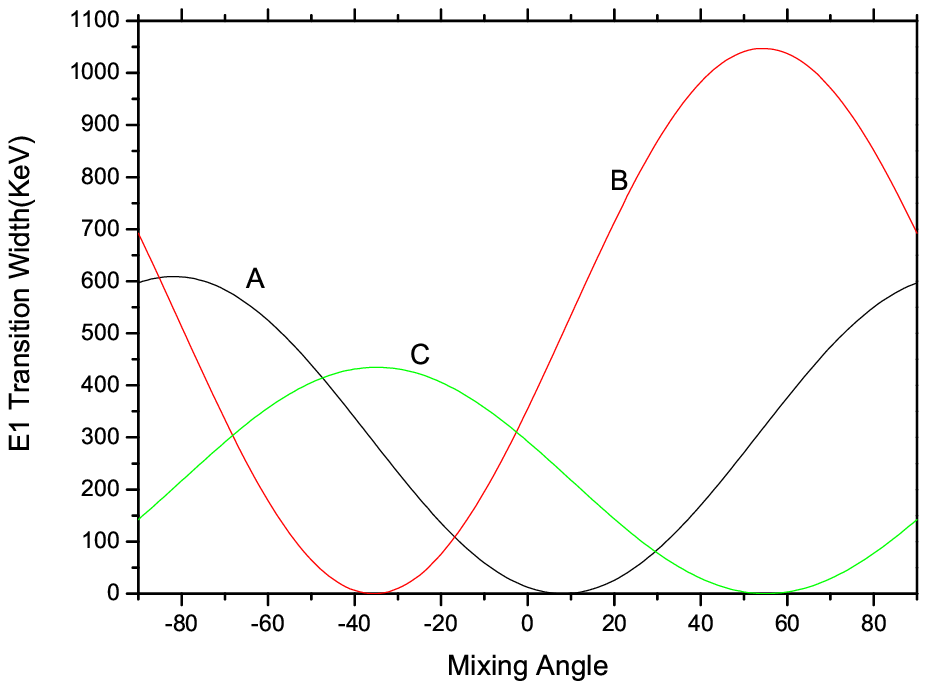}
\caption{E1 transition widths of $\psi(4160)$ varying with 3S-2D
mixing angle in Model I. A: $\psi(4160) \to
\chi^{'}_{c_2}(3929)\gamma$, B: $\psi(4160) \to
\chi^{'}_{c_1}(3872)\gamma$, C: $\psi(4160) \to
\chi^{'}_{c_0}(3915)\gamma$.} \label{ModIpsi4160to2P}
\end{figure}

\begin{figure}
 \centering
\includegraphics[width=10cm,height=7.5cm]{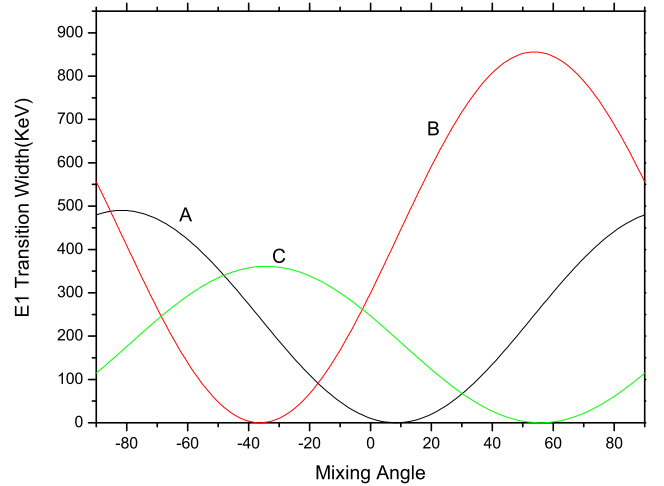}
\caption {E1 transition widths of $\psi(4160)$ varying with 3S-2D
mixing angle in Model II. A: $\psi(4160) \to
\chi^{'}_{c_2}(3929)\gamma$, B: $\psi(4160) \to
\chi^{'}_{c_1}(3872)\gamma$, C: $\psi(4160) \to
\chi^{'}_{c_0}(3915)\gamma$.} \label{ModIIpsi4160to2P}
\end{figure}

The corresponding E1 transition widths parameterized by the mixing angle are
\begin{widetext}
\begin{eqnarray}
\Gamma(\psi(4040)\rightarrow\chi^{'}_{c_0}\gamma)&=&\frac{4}{27}\alpha
e_{c}^2k^3(cos^2\theta <2^3P_0|r|3^3S_1>^2-2\sqrt{2}cos\theta
sin\theta <2^3P_0|r|3^3S_1><2^3P_0|r|2^3D_1>
\nonumber\\&&+2sin^2\theta
<2^3P_0|r|2^3D_1>^2),\\
\rule{0.0cm}{0.7cm}\Gamma(\psi(4040)\rightarrow\chi^{'}_{c_1}\gamma)&=&\frac{4}{9}\alpha
e_{c}^2k^3(cos^2\theta <2^3P_1|r|3^3S_1>^2+\sqrt{2}cos\theta
sin\theta <2^3P_1|r|3^3S_1><2^3P_1|r|2^3D_1>
\nonumber\\&&+\frac{1}{2}sin^2\theta
<2^3P_1|r|2^3D_1>^2),\\
\rule{0.0cm}{0.7cm}\Gamma(\psi(4040)\rightarrow\chi^{'}_{c_2}\gamma)&=&\frac{20}{27}\alpha
e_{c}^2k^3(cos^2\theta
<2^3P_2|r|3^3S_1>^2-\frac{\sqrt{2}}{5}cos\theta sin\theta
<2^3P_2|r|3^3S_1><2^3P_2|r|2^3D_1>
\nonumber\\&&+\frac{1}{50}sin^2\theta
<2^3P_2|r|2^3D_1>^2),\\
\rule{0.0cm}{0.7cm}\Gamma(\psi(4160)\rightarrow\chi^{'}_{c_0}\gamma)&=&\frac{4}{27}\alpha
e_{c}^2k^3(2cos^2\theta <2^3P_0|r|2^3D_1>^2+2\sqrt{2}cos\theta
sin\theta <2^3P_0|r|2^3D_1><2^3P_0|r|3^3S_1>
\nonumber\\&&+sin^2\theta <2^3P_0|r|3^3S_1>^2,\\
\rule{0.0cm}{0.7cm}\Gamma(\psi(4160)\rightarrow\chi^{'}_{c_1}\gamma)&=&\frac{4}{9}\alpha
e_{c}^2k^3(\frac{1}{2}cos^2\theta
<2^3P_1|r|2^3D_1>^2-\sqrt{2}cos\theta sin\theta
<2^3P_1|r|2^3D_1><2^3P_1|r|3^3S_1>  \nonumber\\
&&+sin^2\theta <2^3P_1|r|3^3S_1>^2),\\
\rule{0.0cm}{0.7cm}\Gamma(\psi(4160)\rightarrow\chi^{'}_{c_2}\gamma)&=&\frac{20}{27}\alpha
e_{c}^2k^3(\frac{1}{50}cos^2\theta
<2^3P_2|r|2^3D_1>^2+\frac{\sqrt{2}}{5}cos\theta sin\theta
<2^3P_2|r|2^3D_1><2^3P_2|r|3^3S_1> \nonumber\\&&+sin^2\theta
<2^3P_2|r|3^3S_1>^2),
\end{eqnarray}
\end{widetext}
We use the lowest-order wave functions calculated in model I-III and
impose $m_{\chi^{'}_{c_2}}=3929\, \mbox{MeV}$,
$m_{\chi^{'}_{c_1}}=3872\, \mbox{MeV}$ and
$m_{\chi^{'}_{c_0}}=3915\, \mbox{MeV}$. The results are similar in
the three models, i.e., the E1 transition widths reach their maximum
or minimum values almost at the same mixing angle in the three
models. As an example, we display the results in Model I and Model
II in Figs.~\ref{ModIpsi4040to2P}-\ref{ModIIpsi4160to2P}, and do not
show the similar results in Model III for simplicity.


We see that the decay width of $\psi(4040) \to
\chi^{'}_{c_2}(3929)\gamma$ reaches its maximum of about 75 KeV
corresponding to a branching ratio of $9.3\times 10^{-4}$ near
$\theta=10^o$. While the decay width of $\psi(4160) \to
\chi^{'}_{c_2}(3929)\gamma$ reaches its minimum (zero) near $10^o$
and its maximum of about 600 KeV near $-80^o$ at which $\psi(4160)$
is almost a pure $\psi(3 ^3S_1)$ and the corresponding branching
ratio is about $6\times 10^{-3}$.

The decay width of $\psi(4040) \to \chi^{'}_{c_1}(3872)\gamma$
reaches its maximum of about 250 KeV  corresponding to a branching
ratio of $3.1\times 10^{-3}$ near $-35^o$, which is about 2 times
larger than that of non-mixing,  and reaches its minimum (zero) near
$55^o$. While the decay width of $\psi(4160) \to
\chi^{'}_{c_1}(3872)\gamma$ reaches its minimum (zero) near $-35^o$
and reaches its maximum of about 1050 KeV corresponding to a
branching ratio of about 1\% near $55^o$. It is very interesting to
note that the two values happen to be corresponding to the mixing
angles determined by leptonic decay widths of $\psi(4040)$ and
$\psi(4160)$. That means, one of the two channels must be enhanced
by the 3S-2D mixing if the mixing mechanism just affect the leptonic
decays and the E1 transitions by simply mixing the wave functions.

The decay width of $\psi(4040) \to \chi^{'}_{c_0}(3915)\gamma$
reaches its maximum of about 52 KeV  corresponding to a branching
ratio of $6.5\times 10^{-4}$ near $55^o$, which is about 2.5 times
larger than that of non-mixing  and reaches its minimum (zero) near
$-35^o$. While the decay width of $\psi(4160) \to
\chi^{'}_{c_0}(3915)\gamma$ reaches its minimum (zero) near $55^o$
and reaches its maximum of about 450 KeV corresponding to a
branching ratio of about $4.5\times 10^{-3}$ near $-35^o$.

Since the decay width of $\psi(4160) \to \chi^{'}_{c_2}(3929)\gamma$
reaches its minimum (zero) and $\psi(4040) \to
\chi^{'}_{c_2}(3929)\gamma$ reaches its maximum near $10^0$, which
is not far from the non-mixing case, in our model calculations, we
may use these two channels to check whether there is substantial
3S-2D mixing between $\psi(4040)$ and $\psi(4160)$. If $
\chi^{'}_{c_2}(3929)$ can be detected in the E1 transitions from
$\psi(4040)$ but not from $\psi(4160)$, then the mixing angle should
be small.  In general, since the Z(3930) is identified with
$\chi^{'}_{c_2}$, the observed E1 transition rates to Z(3930) from
$\psi(4040)$ and $\psi(4160)$ will be useful to constrain the value
of 3S-2D mixing angle by comparing the measurements and the
theoretical predictions. When we have a better control over the
value of mixing angle, we will be in a position to study the
properties of X(3872) and X(3915), and to see if they can be
respectively the $\chi^{'}_{c_1}$ (or partially be) and
$\chi^{'}_{c_0}$  by comparing the observed E1 transition rates with
theoretical predictions for $\psi(4040)$ and $\psi(4160)$ decays to
$\chi^{'}_{c_{1,0}}\gamma$. However, we must keep in mind that the
3S-2D mixing model for $\psi(4040)$ and $\psi(4160)$ is only a
simplification for the real situation, since the mixing with 4S
state and the charm meson pairs are all neglected in the 3S-2D
mixing model. Nevertheless, with this simple model we hope some
useful information can be obtained for these X,Y,Z states by
measuring the E1 transition rates of $\psi(4040)$ and $\psi(4160)$
to $\chi^{'}_{c_J}$ (J=2,1,0) charmonium states.


We also calculate the E1 transition widths of
$\psi(4040,4160)\rightarrow \chi_{c_J}(1P)\gamma$ in order to see
whether these can be helpful in determining the 3S-2D mixing angle.
The results are listed in Table~\ref{E1rad3}. Although the
calculated widths are expectedly small, the transition branching
ratio of $\chi_{c_{1,2}}\rightarrow J/\psi\,\gamma$ are relatively
large (about 20\% for $\chi_{c_2}$ and 36\% for $\chi_{c_1}$). So
hopefully it is possible to measure them in experiment if the data
samples are large enough. A special case is $\psi(4160)\rightarrow
\chi_{c_1}\gamma$ with a branching ratio of order $10^{-4}$, which
is quite large, and might be easier to be detected. Again, if we
have determined the mixing angle, it would help us searching for
$\chi^{'}_{c_J}$ and identify the X,Y,Z resonances.

\begin{table*}
\caption{E1 transition widths and branching ratios of charmonium
states in various potential models with lowest-order wave functions.
The masses and total widths of the initial and final states used in
the calculation  are the PDG~\cite{Nakamura:2010zzi} central
values.} \vskip 0.3cm
\begin{ruledtabular}
\begin{tabular}{ll|lll|lll|lllll|lllll }
\multicolumn{2}{c|}{Process}          &\multicolumn{3}{c|}{$<f|r|i>\,(GeV^{-1}$)} &\multicolumn{3}{c|}{$\rm k\,(MeV)$}                     & \multicolumn{5}{c|}{$\Gamma_{\rm thy}$~(keV)}                                     & \multicolumn{5}{c}{$Br_{\rm thy}~(\times 10^{-4})$}                                  \\
     Intial  &Final                    &I     &II    &III                          &Ours  &Ref.\cite{Barnes2005}&Ref.\cite{Godfrey:1985xj}     &I         &II      &III     &Ref.\cite{Barnes2005}&Ref.\cite{Godfrey:1985xj}       &I   &II   &III  &Ref.\cite{Barnes2005}&Ref.\cite{Godfrey:1985xj}      \\
\hline
 $\psi(4040)$&$\chi_{C2}(3556)$        &0.026 &0.078 &-0.011                       &454  &455                 &508                          &0.16      &1.4     &0.03    &0.70                &12.7                           &0.02&0.18 &0.004&0.09                &1.6\\
             &$\chi_{C1}(3511)$        &0.026 &0.078 &-0.011                       &493  &494                 &547                          &0.12      &1.1     &0.02    &0.53                &0.85                           &0.02&0.14 &0.003&0.07                &0.11\\
             &$\chi_{C0}(3415)$        &0.026 &0.078 &-0.011                       &576  &577                 &628                          &0.06      &0.6     &0.01    &0.27                &0.63                           &0.01&0.08 &0.001&0.03                &0.08\\

\hline
 $\psi(4160)$&$\chi_{C2}(3556)$        &0.43  &0.34  &0.23                         &554  &559                 &590                          &1.5       &0.9     &0.4     &0.79                &0.027                          &0.15&0.09 &0.04 &0.08                &0.003  \\
             &$\chi_{C1}(3511)$        &0.43  &0.34  &0.23                         &592  &598                 &628                          &28        &17      &7.9     &14                  &3.4                            &2.7 &1.7  &0.77 &1.4                 &0.33\\
             &$\chi_{C0}(3415)$        &0.43  &0.34  &0.23                         &672  &677                 &707                          &55        &33      &15      &27                  &35                             &5.3 &3.2  &1.5  &2.6                 &3.4 \\
 \end{tabular}
\end{ruledtabular}
\label{E1rad3}
\end{table*}

\section{summary}
In this paper, we calculate the E1 transition widths and branching
ratios of $\psi(4040,4160,4415)\rightarrow \chi^{'}_{c_J}\gamma$ in
three typical potential models with both lowest- and first-order
relativistically corrected wave functions. We find the transition
widths of $\psi(4040,4160)\rightarrow \chi^{'}_{c_J}$ are
model-insensitive and relatively large (tens to hundreds of KeV) and
the corresponding branching ratios are of order $10^{-4}-10^{-3}$,
which make the search for $\chi^{'}_{c_J}$ possible at $e^+e^-$
colliders such as BEPCII/BESIII and CESR/CLEO. This may help us
identify some of the recently discovered X,Y,Z particles, especially
the $X(3915)$, $Y(3915)$, $Y(3940)$ and $X(3872)$. The possible
3S-2D mixing of $\psi(4040)$ and $\psi(4160)$ and its effect on the
transition widths of $\psi(4040,4160)\rightarrow
\chi^{'}_{c_J}\gamma$ are considered and found to be important. We
find the transitions $\psi(4040,4160)\rightarrow
\chi^{'}_{c_2}\gamma$ can be used to examine whether the mixing
exists and to further possibly constrain the mixing angle; and the
transitions $\psi(4040,4160)\rightarrow \chi^{'}_{c_1}\gamma$ and
$\psi(4040,4160)\rightarrow \chi^{'}_{c_0}\gamma$ can be used to
estimate how far is the mixing angle from $-35^o$ and $55^o$, which
are determined in a simple 3S-2D mixing model by the observed
leptonic decay widths of $\psi(4040)$ and $\psi(4160)$. The
transitions of $\psi(4040,4160)\rightarrow \chi_{c_J}(1P)\gamma$ are
also discussed, and the $\psi(4160)\rightarrow \chi_{c_1}\gamma$
transition is found to have a branching ratio of order of $10^{-4}$,
which may be relatively easier to be detected.

\section{acknowledgement}
We would like to thank Chang-Zheng Yuan for many valuable
discussions. This work was supported in part by the National Natural
Science Foundation of China (No 11047156, No 11075002, No 11021092,
No 10905001), and the Ministry of Science and Technology of China
(2009CB825200).

\end{document}